\documentstyle[prb,aps,psfig]{revtex}
\begin{document}
\newcommand {\be}{\begin{equation}}
\newcommand {\ee}{\end{equation}}
\newcommand {\bea}{\begin{eqnarray}}
\newcommand {\eea}{\end{eqnarray}}
\newcommand {\nn}{\nonumber}

\draft
%
%\twocolumn[\hsize\textwidth\columnwidth\hsize\csname @twocolumnfalse\endcsname
%
%
% Title Page
%

\title{
Excitation Spectra 
and Thermodynamic Response of Segmented Heisenberg Spin Chains
}

\author{Stefan Wessel and Stephan Haas}
\address{Department of Physics and Astronomy, University of Southern
California, Los Angeles, CA 90089-0484}

\date{\today}
\maketitle

\begin{abstract}
The spectral and thermodynamic response of segmented quantum spin chains
is analyzed using a combination of numerical techniques and finite-size
scaling arguments. Various distributions of segment lengths 
are considered,
including the two extreme cases of quenched and annealed averages. As the
impurity concentration is increased, it is found that (i) the integrated
spectral weight
is rapidly reduced, (ii) a pseudo-gap feature opens up at small frequencies,
and (iii) at larger frequencies a discrete peak structure emerges, dominated
by the contributions of the smallest cluster segments. The corresponding
low-temperature thermodynamic response has a divergent contribution due to 
the odd-site clusters and a sub-dominant exponentially
activated component due to the even-site segments whose finite-size gap is
responsible for the spectral weight suppression at small frequencies.
 Based on simple scaling arguments, 
approximate low-temperature expressions are derived for the uniform 
susceptibility and the heat capacity. These are shown to be in good
agreement with
numerical solutions of the Bethe ansatz equations for ensembles of open-end
chains. 
\end{abstract}
\pacs{}
%\vskip2pc]

\section{Introduction}

Low-dimensional electron systems are known to be particularly sensitive to 
disorder\cite{abrahams}. 
It is therefore difficult to  
realize pure one- or two-dimensional 
behavior at very low temperatures and small frequencies in nature.\cite{footnote1} 
This is somewhat 
disappointing in light of recent detailed theoretical
predictions for the low-energy 
scaling behavior of paradigmatic quantum-many-body models, such as 
antiferromagnetic spin chains and ladders
\cite{eggert,troyer,dagotto,rice,damle}.
It is the enhanced quantum fluctuations in these compounds,
which give rise to particular low-temperature scaling regimes in very pure
samples, and at the same instance
make them highly unstable towards externally induced
low-temperature transitions, 
such as localization by impurity scattering
or three-dimensional long-range
ordering due to residual small couplings between the lower-dimensional
subsystems\cite{wessel,giamarchi}.

It has recently been demonstrated that in certain 1D subsystems,
such as random-exchange and random-spin chains, anomalously 
extended states can persist against disorder
\cite{doty,haas,roder,westerberg,brouwer}. 
The physical picture is that while most spins are bound in randomly
distributed 
valence bonds, the unbound spins interact via virtual 
excitations, resulting in a zero-frequency band with power-law scaling.
\cite{footnote2}
Furthermore,
in the case of spin ladders and spin-Peierls compounds,
doping with randomly placed non-magnetic impurities may actually induce 
quasi-long-range ordering due to effective
inverse-power-law interactions between
the residual ``pruned" spins\cite{sigrist2,laukamp}.
This replacement of an  originally short-ranged spin liquid state 
by impurity-induced quasi-long-range order
can be viewed as a realization of the ``order by disorder"
phenomenon. 

In other compounds, impurity scattering may completely destroy
the connectivity within
one-dimensional subsystems. If this is the case
extended states cannot survive. 
Let us examine two specific realizations of such segmented quantum
spin systems: (i) CuO chains, with non-magnetic 
impurities, and (ii) pinned charge density waves in quasi-1D 
materials.
The first situation can be realized by doping 
a quasi-1D compound such as SrCuO$_2$ with Zn.
In the pure material, antiferromagnetic superexchange between 
neighboring Cu$^{2+}$
$\rm d_{x^2-y^2}$ electrons is mediated by the filled O$^{2-}$ p-orbitals.
By substituting Zn$^{2+}$ for Cu$^{2+}$, static vacancies are created,
and the infinite chain is separated into segments of length $l$ which follow
a discrete distribution.\cite{haas2,yamazaki}
A second physical way of realizing segmented spin chains is the pinning 
of one-dimensional charge density waves. If there is competition
between poorly screened long-ranged repulsive Coulomb forces and
short-range attractive forces, highly inhomogeneous density wave modulations
occur, favoring particular segment lengths.\cite{hiroi}
Even small impurity scattering leads to a pinning of 
such structures. 

In both cases, 
there are ensembles of correlated spin segments which 
are most straightforwardly modelled by taking appropriate averages
over distributions of finite clusters with open boundary conditions. The 
specific form 
of the distribution function strongly depends on the details of how the
segments are formed. For example, in the case of randomly doped CuO chains
a (discrete) Poisson distribution is natural\cite{haas2,yamazaki}, 
whereas for pinned, spatially
modulated density waves, only two or three cluster sizes may dominate.
An important factor, determining the proper distribution function, is the 
(meta-)stability of the random realizations: 
are they obtained from a quenched or an annealed cooling procedure? While
in quenched realizations all clusters that occur at high
temperatures also
have non-vanishing weight in the zero-temperature distribution
function, slower ``annealed" cooling processes can
lead to preferred sizes and shapes. In particular, segments with an electronic
closed shell configuration have more stable groundstates than others, and 
thus receive a higher weight in an annealed cooling process.\cite{footnote3}    

In this work, 
we systematically study
such ensembles of antiferromagnetically correlated spin clusters,
focusing
on the evolution of the corresponding
low-energy features in the dynamical
spin excitation spectrum 
and on the uniform static susceptibility
as a function of the hole concentration.
In particular for small clusters, quantum effects are important 
\cite{footnote3}, which makes any theoretical approach to this problem
challenging. We therefore attack this task numerically,
using exact numerical diagonalization \cite{lanczos}
and scaling laws derived from conformal field theory
to calculate the static and dynamical magnetic response
for variable
segment sizes\cite{haas2}.

This paper is organized as follows. In the next section, a procedure
to obtain excitation spectra for segmented Heisenberg chains is 
explained. The appropriate distribution functions are derived, and the 
evolution of the spectra with impurity concentration is discussed. 
In the subsequent section, the static magnetic response, i.e. the uniform
susceptibility, and the heat capacity are
calculated for various ensembles of finite chains.
In the final section we conclude with a discussion of possible extensions
and experimental
consequences of the procedure outlined in this paper.

\section{Excitation Spectra at T=0}

\begin{figure}
\centerline{\psfig{figure=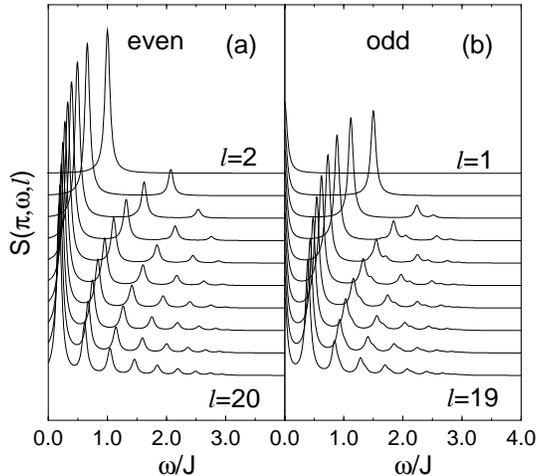,width=7cm,height=6.0cm,angle=270}}
\vspace{0.5cm}
\caption{Finite-size scaling of the staggered dynamical structure factor of the antiferromagnetic spin-$\frac{1}{2}$ Heisenberg chain with open boundary conditions. The poles have been given a width of $\epsilon=0.1 J$.
(a) chains with an even number of sites, 
(b) chains with an odd number of sites. }
\end{figure}

To study the zero-temperature
dynamical response of randomly segmented Heisenberg chains we 
examine the dynamical structure factor $S(q,\omega,l)$ 
of finite chains with open boundary conditions:
\begin{equation}
S({\bf q},\omega,l)=
\frac{1}{l Z}\sum_{n=1}^l |<n|S^z_{{\bf q}}|0>|^2 \delta(\omega-E_n+E_0),
\end{equation}
where $l$ is the length of the finite chain, $Z$ the partition function 
and  $n$ runs over all possible final states, whereas $|0>$ is 
the groundstate.
Let us first concentrate on the staggered magnetization
\begin{equation}
S^z_{\pi}=\sum_{n=1}^l (-1)^n S^z_n.
\end{equation}
Then $S(\pi,\omega,l)$
is well defined for open boundary conditions and chains of an even or 
odd number of sites $l$. In the case of even $l$ there is a unique 
singlet groundstate, whereas one has to take into account the doublet 
nature of the groundstate for odd $l$.
Using numerical diagonalization techniques, we have obtained
$S(\pi,\omega,l)$ with $l=1, ..., 20$. The resulting spectra are shown in Fig.
1.
We observe that segments of even length $l$ have $l/2$ major peaks at non-zero 
frequencies, whereas those of odd length have a significant pole
at $\omega=0$ and $(l-1)/2$ additional peaks at higher frequencies.  
The higher energy peaks have a more complex structure, because some of the
final states are singlets. Zero-frequency peaks occur only in the
odd length segments, reflecting the fact that their groundstates 
are doublets.

 As in the case of closed finite chains \cite{haas2} 
equations derived from conformal field theory can be used
to extract the finite-size 
scaling behavior of $S(\pi,\omega,l)$.\cite{affleck,alcaraz,nomura,logarithmic}
To leading order, 
the pole positions and amplitudes are given by \cite{haas2,logarithmic} 
\begin{eqnarray}\label{cftfitt}
\omega_i(l) & = & \alpha_i/l + \beta_i/l[\ln{(l)}],\\
A_i(l) & = & a_i + b_i\ln{(1+l)} + c_i \ln{[1 + \ln{(1+l)}]},
\end{eqnarray}
where the coefficients $\alpha_i, \beta_i, a_i, b_i,$ and $c_i$ can be
treated as
fit parameters. In Fig. 2 the pole positions and amplitudes of the lowest
three poles are shown
for clusters of up to 20 sites, along with the fits to the above scaling 
equations. In particular for the larger-size segments, these equations give
an excellent fit to the numerical data. In the shorter segments 
higher-order logarithmic corrections for the amplitudes become increasingly 
relevant, and the quality of the fits deteriorates slightly in this 
regime. Note also, that the peak of the one-site cluster at $\omega=0$ 
has an amplitude that does not follow the general trend. 
The amplitudes shown in Fig. 2 correspond to the dynamical spin
response per site (Eq. (1)). 
In the following, we will consider averages over ensembles of finite chains, 
where the amplitudes of the individual segments enter as extensive quantities.
In this
case, the segment amplitudes per site shown in Fig. 2 have to be 
multiplied by the segment length l (Eqs. 5 and 6).

\begin{figure}
\centerline{\psfig{figure=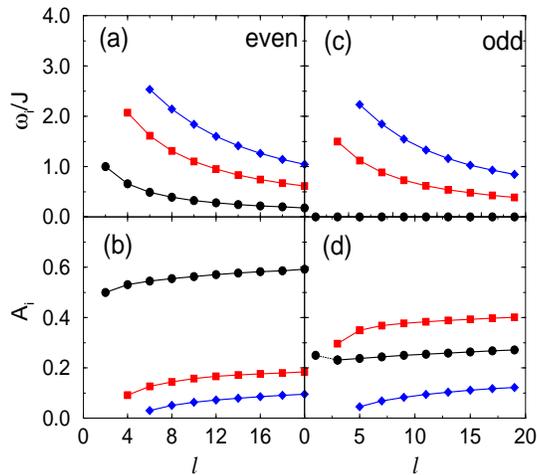,width=7cm,height=6.0cm,angle=270}}
\vspace{0.5cm}
\caption{Finite-size scaling of the first three poles of the staggered 
dynamical structure factor of spin-$\frac{1}{2}$ Heisenberg chains with 
open boundary conditions. The symbols show the values obtained by
numerical diagonalizations, and the solid lines are the result of a fit to 
equations (3) and (4). Circles: first poles, squares: second poles, 
diamonds: third poles. (a) pole positions for chains of even length, 
(b) the corresponding amplitudes, (b) and (d) analogous
results for chains of odd length.}
\end{figure}
Now consider an infinite Heisenberg chain, doped with non-magnetic 
impurities, resulting in an ensemble of open-end segments 
of various lenghts l.
The average dynamical spin structure factor can then be calculated
from
\begin{eqnarray}\label{wholesum0}
S(\pi,\omega) = \sum_l l P(l) S(\pi,\omega,l),
\end{eqnarray}
where $P(l)$ is an appropriately chosen distribution function.
Using the pole structure of the response functions for the individual segments, we obtain
\begin{eqnarray}\label{wholesum}
S(\pi,\omega) = \sum_l \sum_i l P(l) A_i(l) \delta(\omega - \omega_i(l)).
\end{eqnarray}
In practice, the $\delta$-function in Eq. 6 is replaced by a
Lorentzian of width 
$\epsilon$, which will be taken as $\epsilon=0.1J$ throughout the paper. 
$P(l)$ determines the weight of each segment in the ensemble
average, and extrinsic factors favoring certain cluster shapes over others
enter through this function. If the chain
segmentation occurs completely randomly, the corresponding  distribution
function is given by $P(l) = \rho^2 (1 - \rho)^l$, where
$\rho \in [0,1]$ is the concentration of vacant sites\cite{yamazaki,stauffer}.
This distribution function is normalized by
\begin{equation}\label{distnorm}
\sum_{l=1}^{\infty} l P(l) =1-\rho,
\end{equation}
and for $\rho\ll 1$ it can be approximated by 
$P(l)\approx \rho^2 \exp(-\rho l)$.
We also note that since the total number of clusters per site is given 
by $n_c=\sum_l P(l)=\rho(1-\rho)$, the average length of the clusters 
is $l_{av}=\sum_l l P(l)/n_c=1/\rho$. The above distribution describes the 
case of quenched disorder in the infinite chain, i.e. the positions of the 
impurities are uncorrelated.  In an annealed cooling process, even length 
segments are favoured over odd ones, because they have a lower groundstate 
energy. We describe this situation by a similar distribution 
function $P_a(l)=C \rho^2 (1-\rho)^l \delta_{0, l\: mod\: 2}$ where C is 
determined by the normalization condition (\ref{distnorm}). Only even 
length spin segments occur in a chain with this distribution function 
for the impurities. Note, that according to common terminology all of the
ensemble averages which are discussed here
are ``quenched" because frozen disorder
realizations are used. In this paper, 
the terms ``annealed" and ``quenched" only refer 
to the cooling procedures giving rise to different 
distribution functions. For true annealed disorder, however, the 
disorder variables would have to be treated dynamically.

Here we wish to evaluate and analyze $S(\pi,\omega)$ and $S(\pi)$
by using the finite-size scaling 
behavior of the lowest few poles in the spectrum. This few-mode approximation 
has been shown to be valid for sufficiently large impurity concentrations,
but it tends to underestimate the spectral response for very small
$\rho$.\cite{haas2} Let us use the lowest three poles of each segment
as shown in Fig. 2,  
i.e. the index $i$ 
in Eq. (\ref{wholesum}) runs from $i=1$ to 3. Also, a cutoff 
length $l_{max}=10 000$ is used in the sum over $l$. 
The resulting 
frequency-integrated staggered dynamical structure factor 
$S(\pi)=\int d\omega S(\pi,\omega)$ is then given by
\begin{equation}
S(\pi)=\sum_{l=1}^{\l_{\max}}l P(l) S(\pi,l),
\end{equation}
where $S(\pi,l)$ is the frequency-integrated dynamical structure 
factor at wavevector $\pi$, with the scaling form
\begin{equation}\label{spifitt}
S(\pi,l)=a+b \ln(1+l)+c \ln[1+\ln(1+l)].
\end{equation}
Comparing the values of $S(\pi)$ obtained from exact numerical diagonalizations
of ensembles of finite clusters with 
the result for $S(\pi)$ within the three-mode approximation, one finds that 
the three-mode approximation tends to underestimate $S(\pi)$ for small
impurity concentrations. The difference in 
the integrated weight is due to neglecting the higher frequency poles 
that become more relevant for larger clusters and thus smaller impurity 
concentrations. This weight can be approximately restored 
by adding the properly normalized dynamical 
structure factor of an
infinite Heisenberg chain, broadend by $\epsilon$ 
to $\Delta S(\pi,\omega)\propto \epsilon/(\omega^2+\epsilon^2)/\pi$, 
neglecting logarithmic corrections. It turns out that these corrections 
to $S(\pi)$ are only relevant for very low impurity concentrations and are 
neglegible for $\rho>0.2$. The resulting spectra are shown for 
different $\rho$ in Fig. 3 for the quenched and the annealed case. 
\begin{figure}
\centerline{\psfig{figure=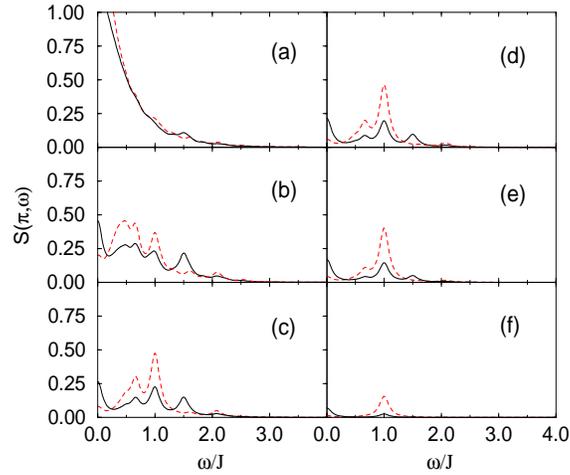,width=7cm,height=6.0cm,angle=270}}
\vspace{0.5cm}
\caption{Evolution of $S(\pi,\omega)$ of the segmented 
antiferromagnetic spin-$\frac{1}{2}$ Heisenberg chain as a function of 
the impurity concentration. (a)-(f) spectra 
for $\rho=0.1 ,0.3, 0.5, 0.6, 0.7, 0.9$. The solid lines 
represent the quenched case, and the dashed lines represent the annealed case.}
\end{figure}
For both types of
 impurity distribution functions there are two main features 
that occur in $S(\pi,\omega)$ upon increasing the number of vacancies in 
the infinite chain. First, the integrated spectral weight decreases rapidly 
upon increasing the impurity concentration. This is shown explicitly 
in Fig. 4, where the solid lines in part (a) and (b) represent the dependence 
of $S(\pi)$ on $\rho$. Note that in the annealed case the integrated 
spectral weight is decreasing at a
slower rate than in the quenched case, especially 
for large impurity concentrations.
 This is due to the fact, that the integrated spectral weight of finite 
chains is a monotonous function of $l$. Thus the total response  increases 
if at constant impurity concentration the odd-length 
segments (starting with $l=1$) are substituted by even-length 
segments (starting with $l=2$).
 Most of the suppressed spectral weight comes from the low-energy continuum of 
the response function due to the large segments. It can be shown 
by considering a lowest-order single-mode approximation that $S(\pi,\omega)$ 
is in fact exponentially suppressed at low frequencies \cite{haas2}.

The second common feature is the emergence of a discrete peak structure 
at larger frequencies (of order J), dominated by the contributions of 
the smallest cluster segments. The dominant segments, occuring according to 
the distribution function $P(l)$, can be identified from these
higher energy peaks. In the annealed case,  
the smallest segments are the two-site clusters with a  pole 
in $S(\pi,\omega,l)$ at $\omega=1J$, and the four-site chain  with a 
major pole at $\omega=0.66J$.
These poles are well separated from the 
low-energy continuum and carry most of the spectral weight of $S(\pi,\omega)$  
at impurity concentrations $\rho\ge 0.5$ (Fig. 3 (c) - (e)).
The pole of the smallest segment 
at $\omega=1J$ dominates the dynamical response function  in the annealed 
case at high concentrations, $\rho>0.7$. One thus expects that 
most of the spectral weight can be obtained from the two-site cluster 
for $\rho$ close to one. In fact, from Eq.(8) one finds
\begin{equation}
S(\pi)\approx2\;P(2)\;S(\pi,2)\approx \frac{1-\rho}{2}, \quad \rho\rightarrow 1,
\end{equation}
explaining the linear behavior of $S(\pi)$ for large impurity concentrations.
 Since in the quenched case the single spin sites and the three-site chains 
are also present, their low frequency poles  contribute strongly to the 
spectral weight. The major pole of the three site chain at $\omega=1.5J$ is 
the first well-defined pole to separate from the low-energy continuum upon 
increasing the impurity concentration, as can be seen in Fig. 3 (a) already 
at $\rho=0.1$. It dominates the spectrum away from the low-energy 
pseudo-branch until, at intermediate concentrations, the lowest-energy poles 
of the two- and four-site clusters are also separated from the continuum.  
Characteristic features of the underlying impurity distribution can thus 
be read off from the higher frequency part of the spectrum.

The major difference between the two impurity distributions we have 
studied lies in 
the rate at which $S(\pi,\omega)$ is suppressed for small frequencies. Consider 
the annealed case first. The even-site segments do not have a pole 
at $\omega=0$ due to their inherent finite spin-gap. Thus
 the exponential suppression of 
the contributions from the large segments leads to the development 
of a pseudogap at small frequencies with increasing impurity concentration.
The finite values of $S(\pi,0)$ are 
due to our replacement of the $\delta$-peak in Eq. (6) by Lorentzians. 
This mimics the various broadening mechanisms in 
real materials, such as thermal broadening, scattering from phonons, and 
interactions with out-of-chain impurities. As shown in Fig. 4 (c), 
the residual spectral weight $S(\pi,0)$
is  exponentially suppressed with increasing $\rho$. 
In Fig. 4 (d) the important higher-energy peaks are compared to $S(\pi,0)$, 
clearly indicating the reduction of the zero-frequency weight in the 
annealed case. 

\begin{figure}
\centerline{\psfig{figure=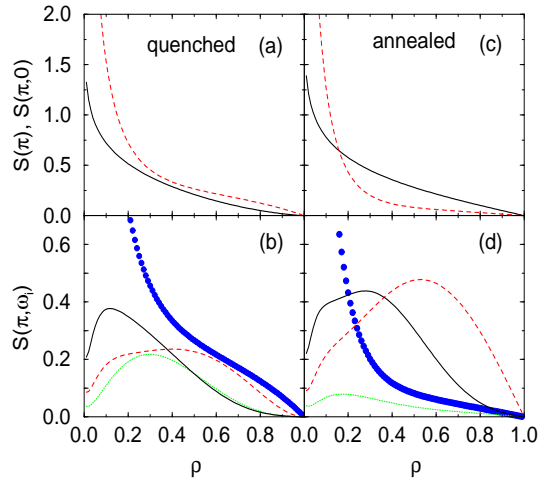,width=7cm,height=6.0cm,angle=270}}
\vspace{0.5cm}
\caption{(a) and (c) The frequency-integrated dynamical 
structure factor at wavevector $\pi$, $S(\pi)$ (solid line) and the 
residual spinon density of states $S(\pi,0)$ (dashed line) as a 
function of the impurity 
concentration $\rho$. (a) quenched disorder, (c) annealed disorder. (b), (d) 
Dependence of the amplitudes of the peaks at $\omega=0$ 
(full circles) $\omega_1=0.66J$ (solid line) , $\omega_2=1.0J$ (dashed line) 
and $\omega_3=1.5J$ (dotted line) on the impurity concentration $\rho$. (c) 
quenched, (d) annealed.} 
\end{figure}

Now consider the quenched case. All odd-length segments 
and especially the dominating one-spin segment show a pole in their dynamical 
response function at zero frequency. Therefore one expects the rate 
of suppression 
of $S(\pi,0)$ to be significantly reduced in the quenched case with respect 
to the annealed case. This is clearly observed in 
Fig. 4, comparing the value of $S(\pi,0)$ for both 
distributions. In the quenched case the weight at $\omega=0$ remains a major 
contribution to $S(\pi)$, even at large impurity concentrations. Fig. 4 (a) 
shows that the reduction of $S(\pi,0)$ with increasing $\rho$
is less pronounced
in the quenched case. In fact,
the amplitude of the peak at $\omega=0$ dominates the spectrum for all $\rho$, 
as shown in Fig 4. (b).

\section{Thermodynamic Response: Uniform Susceptibility and Heat Capacity}
In this section we examine the thermodynamic response of segmented
spin-1/2 Heisenberg chains.  Based on simple scaling arguments,
low-temperature approximations for the uniform 
susceptibility and the specific heat are derived and shown to be in good
agreement with a numerical solution of the problem, using the Bethe ansatz
equations for open-end chains.
The impurity tail of the 
low-temperature susceptibility (i.e. the divergence of $\chi(T)$ at $T=0$) 
in an ensemble of segmented chains is caused by the odd-length segments. 
Before examining
the sub-dominant contributions of the even-length segments it is 
therefore 
necessary to first identify and discuss
the dominant divergent contributions of the odd-length clusters.
This can be achieved by analyzing $[\chi(T) T]$. 
Its value at zero temperature, $\lim_{T\rightarrow 0} [\chi(T)T]$,
 gives the prefactor of the low-temperature impurity tail in $\chi(T)$.
For chains of even length, which always have a finite spin gap, the value of 
$[\chi(T) T]$ approaches $0$ as T goes to zero, whereas for odd 
lengths $l$, $\lim_{T\rightarrow 0} [\chi(T) T]=1/(4l)$. It follows 
that in the quenched case
\begin{equation}
[\chi(T) T]_0(\rho)\equiv \lim_{T\rightarrow 0} [\chi(T) T] (\rho) =\lim_{T\rightarrow 0} \sum_{l=1}^{\infty} l P(l) [\chi(T) T] (l)=\frac{\rho(1-\rho)}{4(2-\rho)},
\end{equation}
with a maximum at $\rho=2-\sqrt{2}\approx 0.586$. 
In the ideal annealed 
case only even-length segments contribute to the average,
and therefore $[\chi(T) T]_0(\rho)$ vanishes. 
The Curie constant is 
$C(\rho)=\lim_{T\to \infty} [\chi(T) T] =(1-\rho)/4 $, 
independent of the length of the contributing segments.

\begin{figure}
\centerline{\psfig{figure=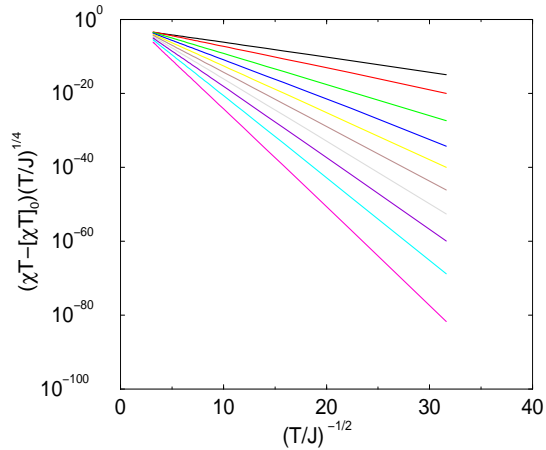,width=7cm,height=6.0cm,angle=270}}
\vspace{0.5cm}
\caption{Linearized form of the scaling behaviour for the low temperature 
susceptibility of segmented Heisenberg chains for impurity 
concentrations $\rho=0.05$ (top) ,0.1,0.2,...,0.9 (bottom). }
\end{figure}

To study the temperature dependence of $\chi(T)$ we can
calculate the finite-size 
susceptibilities from the Bethe ansatz equations which determine the 
energy levels of open Heisenberg chains:
\begin{equation}
\left(\frac{x_k+i}{x_k-i}\right)=\prod_{j\neq k}^{M}\frac{x_k-x_j+2i}{x_k-x_j-2i}\: \frac{x_k+x_j+2i}{x_k+x_j-2i}.
\end{equation}

Here the number of roots M determines the total $S^z$ component of the 
state through the relation $S^z=l/2-M$. In logarithmic form this
equation becomes
\begin{equation}
2l\: \mbox{ctan}^{-1}(x_k)=\pi I_k+ \sum_{j\neq k}^M \left[\mbox{ctan}^{-1}\left(\frac{x_k-x_j}{2}\right)+\mbox{ctan}^{-1}\left(\frac{x_k+x_j}{2}\right)\right],
\end{equation}
where all the $I_k$ are integers with $k=1,...,M$.
Given a solution of the above equations for a set of $I_k$'s, the 
energy of the corresponding eigenstate is
\begin{equation}
\frac{E}{J}=\frac{l-1}{4}-2\sum_{k=1}^M\frac{1}{x_k^2+1}
\end{equation}
The groundstate in a given $S^z$ sector, $E_0(S^z,l)$, is obtained   
from the set $\{I_k\}=\{l+1,l+3,l+5,...,l+2M-1\}$. And the first exited state 
in the $S^z>0$ sectors can be determined from the set 
$\{I_k\}=\{2S^z,2S^z+3,2S^z+5,2S^z+7,...,l-1\}$. 
Using an iterative method to determine the energy gaps for the even 
length segments, $\Delta E_l=E_0(S^z=1,l)-E_0(S^z=0,l)$, one obtains for the 
low temperature susceptibility of the segmented chain:
\begin{equation}
[\chi(T) T](\rho) =[\chi(T) T]_0(\rho) + 2 \sum_{l=2}^{\infty}{}'\:\frac{P(l)}{1+\exp(\Delta E_l/T)},
\end{equation}
where the sum is restricted to even $l$. This equation can be evaluated 
numerically, using the energies from the finite-size Bethe ansatz equations
(Eq. 14) and an appropriate distribution function $P(l)$. It is found that 
the results for $[\chi(T) T]$ typically converge below a length cut-off
around $l_{max} = 1000$. Results for the
quenched case are shown in Fig. 5. For low temperatures 
there appears to be a universal scaling behavior
which can be elucidated by expanding Eq. 15.

To determine the low-temperature scaling behavior of the susceptibility 
we consider the lowest-order finite-size scaling behavior of the energy 
gap, $\Delta E_l\approx \alpha_1/l$, neglecting higher-order
logarithmic corrections
(Eq. \ref{cftfitt}) which become increasingly important when larger 
segments occur (i.e. at very small impurity concentrations).
For low temperatures one obtains
\begin{eqnarray}
[\chi(T) T](\rho) -[\chi(T) T]_0(\rho)&\approx&\:\int_0^{\infty} \exp\left[-\frac{\alpha_1}{T l}-\ln\left(\frac{1}{1-\rho}\right) l\right]\: dl
=\sqrt{\frac{4 \alpha_1}{T\ln\left(\frac{1}{1-\rho}\right)}}\:K_1\left[\sqrt{\frac{4\alpha_1}{T}\ln\left(\frac{1}{1-\rho}\right)}\:\right] \nonumber \\
&\approx&\sqrt{\frac{\pi}{2}\sqrt{\frac{\alpha_1}{T}}}\:\left[ \ln  \frac{1}{1-\rho}\right]^{-\frac{3}{4}}\: \exp\left[-\sqrt{\frac{4\alpha_1}{T} \ln \left(\frac{1}{1-\rho}\right)}\:\right]\:,\quad \quad T\to 0.
\end{eqnarray}
The expected low-temperature behavior is thus of the form
\begin{equation}
[\chi(T) T] (\rho) -[\chi(T) T]_0(\rho)\propto \: \left(\frac{T}{J}\right)^{-\frac{1}{4}}\: e^{-\gamma/\sqrt{T/J}},
\end{equation}
where $\gamma$ turns out to be
\begin{equation}\label{propgamma}
\gamma=\sqrt{4 \;\frac{\alpha_1}{J}\; \ln\left(\frac{1}{1-\rho}\right)}.
\end{equation}

\begin{figure}
\centerline{\psfig{figure=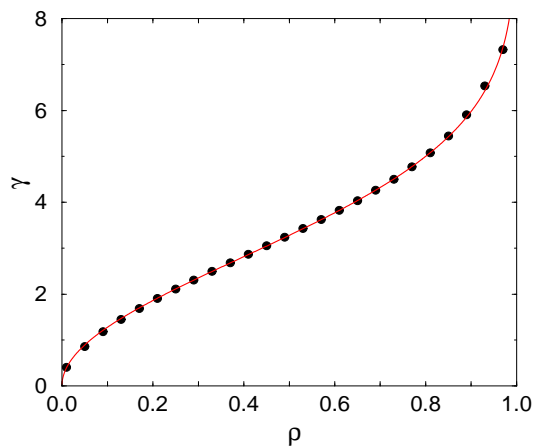,width=7cm,height=6.0cm,angle=270}}
\vspace{0.5cm}
\caption{Fit of the effective low-temperature gap $\gamma$. 
The circles show the numerical solution of the Bethe ansatz equations,
and the solid line is the best 
fit with $\alpha_1=3.88J$. }
\end{figure}

The observed linearity in Fig. 5 where
we have plotted $\ln[(\chi(T) T(\rho) -[\chi(T) T]_0(\rho))(T/J)^{1/4}]$ 
vs. $(T/J)^{-1/2}$ for various $\rho$ confirms the validity of this
low-temperature expansion. The free parameter $\alpha_1$ can be determined
by fitting $\gamma(\rho)$ of Eq. \ref{propgamma} to the exact numerical 
Bethe ansatz solution of $[\chi(T) T](\rho)$ (Eq. 15). 
As shown in Fig. 6, 
for sufficiently small concentrations the fit turns out to be very good,
giving $\alpha_1=3.88J$.

A similar procedure can be used to determine the scaling behavior of the 
specific heat of segmented Heisenberg spin-1/2 chains. For the quenched
case, one finds 
\begin{equation}
C \propto \left(\frac{T}{J}\right)^{-\frac{5}{4}} e^{-\gamma'/\sqrt{T/J}}
\end{equation}
at sufficiently low temperatures. The corresponding effective gap $\gamma'$
turns out to be 
\begin{equation}
\gamma'=\gamma_0+\sqrt{4\;\frac{\alpha_1'}{J}\;\ln\left(\frac{1}{1-\rho}\right)},
\end{equation}
where a fit of the numerical data yields
$\gamma_0=0.182$ and $\alpha'_1=4.106J$.

\section{Conclusions}

In summary, we have examined the spectral and thermodynamic response of 
various ensembles of segmented antiferromagnetic Heisenberg spin-1/2 chains. 
We have calculated the dynamical spin structure factor which 
can be probed by Neutron scattering experiments. The particular dependence 
of this quantity on the impurity concentration is determined by the 
distribution of segment lengths. However, there are several generic features
which are observed for the most common distribution functions. These are
(i) a rapid decay of the integrated spectral weight with increasing impurity
concentration, (ii) a suppression of low-frequency poles (pseudo-gap), 
and (iii) the emergence of a discrete pole structure at higher energies,
dominated by the smallest contributing clusters in the average. While we
find that the first two predictions are consistent with the presently 
available experimental data, point (iii) might be the hardest to verify
because the corresponding signals in a highly disordered sample are typically
rather small and broad.

Two main contributions to the low-temperature 
thermodynamic response of segmented chains can be identified:
 a dominant divergent
component from the odd-side clusters and a sub-dominant exponentially activated
component due to the even-site clusters. The sub-dominant contribution is 
analyzed by subtracting the impurity tail from the total response function,
as it is commonly done in the analysis of
experiments.\cite{yamazaki,azuma} 
Assuming a certain type of distribution function, 
the corresponding effective gaps, here $\gamma$ for the uniform
susceptibility and $\gamma'$ for the specific heat, can be calculated.
An analysis of these gaps can thus be used
as an indicator of the underlying distribution function of segment lengths.

Complete segmentation of quantum spin chains, as it has been treated 
in this work, can be viewed as an extreme case of impurity scattering
with an infinitely large on-site repulsive potential.
While this mechanism  may indeed lead 
to segmentation, in many physical realizations
longer-range exchange paths exist which can partially restore extended
states of the undoped parent systems. Consider for example Zn-doped 
$\rm CuGeO_3$. This compound is known to have sizeable next-nearest neighbor
exchange interactions along the $\rm CuO_2$-chain direction, giving rise to 
an effective $J_1-J_2$ model. Below the transition temperature
$T_{SP}$ the compound goes into a dimerized spin-Peierls phase, whereas above
$T_{SP}$ it is in a critical quasi-one-dimensional state. 
This spin-Peierls transition is suppressed upon replacing the Cu-atoms
randomly with Zn\cite{martin},
most likely because the partial segmentation due to the non-magnetic impurities
impedes the quantum-critical extended states within the chains which in turn
 are a
prerequisite for Peierls transitions. Because of the longer-range 
exchange paths $J_2$, these extended states are not completely destroyed, and a
remnant spin-Peierls phase is observed in 
$\rm Cu_{1-x}Zn_xGeO_3$ at sufficiently
small impurity concentrations x.\cite{martin} 

The segmentation of critical one-dimensional system competes 
with transitions such as three-dimensional ordering due to small
inter-chain interactions or Peierls-type transitions. Segmented 
one-dimensional phases are more stable against low-temperature ordering 
transitions, and critical states can in turn be created in a controlled
manner by impurity-doping short-range ordered systems, for example by
introducing non-magnetic sites into quasi-one-dimensional spin liquids
such as two-leg spin-1/2 Heisenberg ladders.\cite{sigrist2,westerberg,frischmuth}

We thank
A. Honecker and B. Normand
for useful discussions,
and acknowledge
the Zumberge Foundation for financial support.


\begin{references}
\bibitem{footnote1} A notable exception are the quasi-1D Quantum Hall edge
currents which can circumvent surface impurities.
\bibitem{footnote2} This response to disorder scattering is in contrast
to random magnetic fields which generally cause exponential decay in
the correlation functions.
\end{references}
\end{document}